\documentstyle[12pt,aps,epsf]{revtex}
\begin{document}
\tighten
\draft
\title{\bf{Relic Gravitational Waves and Their Detection}}

\author{L. P. Grishchuk\thanks{e-mail: grishchuk@astro.cf.ac.uk}}
\address{Department of Physics and Astronomy, Cardiff University, 
Cardiff CF2 3YB, United Kingdom \\ and \\ Sternberg Astronomical
Institute, Moscow University, Moscow 119899, Russia}

\maketitle
\vspace{2cm}
    
\begin{abstract}
The range of expected amplitudes and spectral slopes
of relic (squeezed) gravitational waves, predicted by theory and 
partially supported by observations, is within the reach of 
sensitive gravity-wave detectors. In the most favorable case, the detection of
relic gravitational waves can be achieved by the cross-correlation of outputs
of the initial laser interferometers
in LIGO, VIRGO, GEO600. In the more realistic case, the sensitivity of 
advanced ground-based and space-based laser interferometers will be needed.  
The specific statistical signature of relic gravitational waves, associated
with the phenomenon of squeezing, is a potential reserve for further 
improvement of the signal to noise ratio.
\end{abstract}


\vspace{1.5cm}

\section{Introduction}  

It is appropriate and timely to discuss the detection of relic 
gravitatational waves at the experimental meeting like this one. We are
in the situation when the advanced laser interferometers, currently under
construction or in a design phase, can make the dream of detecting relic
gravitons a reality. The detection of relic gravitational waves is 
the only way to learn about the evolution of the very early Universe, 
up to the limits of Planck era and Big Bang.
\par
The existence of relic gravitational waves is a consequence of quite
general assumptions. Essentially, we rely only on the validity of general 
relativity and basic principles of quantum field theory. The strong 
variable gravitational field of the early Universe amplifies the inevitable
zero-point quantum oscillations of the gravitational waves and produces
a stochastic background of relic gravitational waves measurable 
today \cite{g1}.
It is important to appreciate the fundamental and unavoidable nature of this
mechanism. Other physical processes can also generate stochastic backgrounds
of gravitational waves. But those processes either involve many additional
hypotheses, which may turn out to be not true, or produce a gravitational wave 
background (like the one from binary stars in the Galaxy) which should 
be treated as an unwanted noise rather than a useful and interesting signal.
The scientific importance of detecting relic gravitational waves has been 
stressed on several occasions (see, for example, \cite{t1}--\cite{sch}). 
\par
The central notion in the theory of relic gravitons is the phenomenon
of superadiabatic (parametric) amplification. The roots of this phenomenon
are known in classical physics, and we will remind its basic features. As
every wave-like process, gravitational waves use the concept of a harmonic
oscillator. The fundamental equation for a free harmonic oscillator is  
\begin{equation}
\label{1}
\ddot{q} + \omega^2 q = 0, 
\end{equation}
where $q$ can be a displacement of a mechanical pendulum or a 
time-dependent amplitude of a mode of the physical field. The energy of the
oscillator can be changed by an acting force or, alternatively, by a 
parametric influence, that is,
when a parameter of the oscillator, for instance the length of a pendulum,
is being changed. In the first case, the fundamental equation takes the
form 
\begin{equation}
\label{2}
\ddot{q} + \omega^2 q = f(t),
\end{equation}
whereas in the second case Eq. (\ref{1}) takes the form
\begin{equation}
\label{3}
\ddot{q} + \omega^2 (t) q = 0.  
\end{equation}
Equations (\ref{2}) and (\ref{3}) are profoundly different, both, 
mathematically and physically. 
\par
Let us concentrate on the parametric influence. We consider a pendulum
of length $L$ oscillating in a constant gravitational field $g$. 
The unperturbed pendulum
oscillates with the constant frequency $\omega = \sqrt{g/L}$. 
$Fig. 1a$ illustrates the 
variation of the length of the pendulum $L(t)$ by an external agent, 
shown by alternating arrows. Since $L(t)$ varies, the frequency 
of the oscillator does also vary: $\omega(t) = \sqrt{g/L(t)}$. 
The variation $L(t)$ does not 
need to be periodic, but cannot be too much slow (adiabatic) if the result
of the process is going to be significant. Otherwise, in the adiabatic
regime of slow variations, the energy of the oscillator $E$ and its 
frequency $\omega$ do change slowly, but $E /\omega$ remains constant, so
one can say that the ``number of quanta" 
$E/ \hbar \omega$ in the oscillator remains fixed.  
In other words, for the creation of new ``particles - excitations", 
the characteristic time of the variation should be 
comparable with the period of the oscillator and the adiabatic behaviour 
should be violated. After some duration of the appropriate parametric influence,
the pendulum will oscillate at the original frequency, but will have a 
significantly larger, than before, amplitude and energy. 
This is shown in $Fig. 1b$. Obviously, the energy of the oscillator has 
been increased at the expense of the external agent (pump field). For
simplicity, we have considered a familiar case, when the
length of the pendulum varies, while the gravitational acceleration $g$
remains constant. Variation of $g$ would represent a gravitational parametric
influence and would be even in a closer analogy with what we study below.    
\par
A classical oscillator must have a non-zero initial amplitude for the
amplification mechanism to work. Otherwise, if
the initial amplitude is zero, the final amplitude will also be zero. 
Indeed, imagine the pendulum strictly at rest, hanging stright down. 
Whatever the variation of its length is, it will not make the pendulum 
to oscillate and gain energy. In contrast,
a quantum oscillator does not need to be excited from the very beginning. 
The oscillator can be
initially in its vacuum quantum-mechanical state. The inevitable zero-point
quantum oscillations are associated with the vacuum state energy
$\frac{1}{2} \hbar \omega$. One can imagine a pendulum hanging stright down,
but fluctuating with a tiny amplitude determined by the ``half of the
quantum in the mode". In the classical picture, it is this tiny amplitude 
of quantum-mechanical origin that is being parametrically amplified.

\begin{figure}
\epsfxsize=0.8\textwidth
\centerline{\epsfbox{oscil.fig1}} 
\end{figure}

\par
The Schrodinger evolution of a quantum oscillator depends crucially 
on whether the oscillator is being excited parametrically or by a force. 
Consider the phase diagram $(q, p)$, where $q$ is the 
displacement and $p$ is the conjugate momentum. The vacuum state is
described by the circle in the center (see $Fig. 2$). The mean values of $q$
and $p$ are zeros, but their variances (zero-point quantum fluctuations) 
are not zeros and are equal to
each other. Their numerical values are represented by the circle in the
center. Under the action of a force, the vacuum state evolves into a coherent 
state. The mean values of $p$ and $q$ have increased, but the variances are 
still equal and are described by the circle of the same size as for the
vacuum state. On the other hand, 
under a parametric influence, the vacuum state evolves into a squeezed
vacuum state. [For a recent review of squeezed states see, for example,
\cite{kn} and references there.] Its variances for the 
conjugate variables $q$ and $p$ 
are significantly unequal and are described by an ellipse. 
As a function of time, the ellipse rotates with respect to the origin 
of the $(q, p)$ diagram, and the numerical values of the variances 
oscillate too. The mean numbers of quanta in the 
two states, one of which is coherent and another is squeezed vacuum,   
can be equal (similar to the coherent and squeezed states shown in $Fig. 2$) 
but the statistical properties of these states are significantly different. 
Among other things, the variance
of the phase of the oscillator in a squeezed vacuum state is very 
small (squeezed). Graphically,
this is reflected in the fact that the ellipse is very thin, so that that
the uncertainty in the angle between the horizontal axis and the orientation 
of the ellipse is very small. This highly elongated ellipse can be
regarded as a portarait of the gravitational wave quantum state that is being 
inevitably generated by parametric amplification, and which we will be 
dealing with below.

\begin{figure}
\epsfxsize=0.8\textwidth
\centerline{\epsfbox{squeeze.fig2}} 
\end{figure}

A wave-field is not a single oscillator, it depends on spatial coordinates 
and time, and may have several independent components (polarization
states). However, the field can be decomposed into a set of 
spatial Fourier harmonics. In this way we represent the gravitational 
wave field as a collection
of many modes, many oscillators. Because of the nonlinear character of
the Einstein equations, each of these oscillators is coupled to the variable
gravitational field of the surrounding Universe. For sufficiently
short gravitational waves of experimental interest, this coupling was
especially effective in the early Universe, when the condition of
the adiabatic behaviour of the oscillator was violated. It is this homogeneous
and isotropic gravitational field of all the matter in the early 
Universe that played the role of the external agent - pump field.  
The variable pump field acts parametrically on  
the gravity-wave oscillators and drives them into multiparticle 
states. Concretely, the initial vacuum state of each pair of waves with
oppositely directed momenta evolves into a highly correlated state 
known as the two-mode squeezed vacuum state \cite{gs}, \cite{g2}. 
The strength and duration of the effective coupling depends 
on the oscillator's frequency. They all start in the vacuum state 
but get excited to various amounts. As a result, a broad 
spectrum of relic gravitational waves is being formed. This spectrum
is accessible to our observations today. 
\par
Let us formulate the problem in more detail.

\section{Cosmological Gravitational Waves}  

In the framework of general relativity, a homogeneous isotropic gravitational
field is decribed by the line element
\begin{equation}
\label{4}
{\rm d}s^2 = c^2{\rm d}t^2 - a^2(t) \delta_{ij} {\rm d}x^i{\rm d}x^j = 
a^2({\eta})[{\rm d}\eta^2 - \delta_{ij} {\rm d}x^i{\rm d}x^j].
\end{equation}
In cosmology, the function $a(t)$ (or $a(\eta)$) is called scale factor.
In our discussion, it will represent gravitational pump field.
\par
Cosmological gravitational waves are small corrections $h_{ij}$ to the metric
tensor. They are defined by the expression 
\begin{equation}
\label{5}
{\rm d}s^2 = a^2({\eta})[{\rm d}\eta^2 - (\delta_{ij} + h_{ij})
{\rm d}x^i{\rm d}x^j].
\end{equation} 
The functions $h_{ij} (\eta ,{\bf x})$ can be expanded over spatial 
Fourier harmonics $e^{i{\bf nx}}$ and $e^{-i{\bf nx}}$, where
${\bf n}$ is a constant wave vector. In this way, we reduce the dynamical
problem to the evolution of time-dependent amplitudes for each 
mode ${\bf n}$. 
Among six functions $h_{ij}$ there are only two independent (polarization) 
components. This decomposition can be made, both, for real and for quantized
field $h_{ij}$. In the quantum version, the functions $h_{ij}$ are 
treated as quantum-mechanical operators. We will use the Heisenberg 
picture in which the time evolution is carried out by the operators while
the quantum state is fixed. This picture is fully equivalent to the
Schrodinger picture, discussed in the Introduction, in which the vacuum
state evolves into a squeezed vacuum state while the operators are time
independent.   
\par
The Heisenberg operator for the quantized real field $h_{ij}$ can 
be written as
\begin{eqnarray}
\label{6}
h_{ij} (\eta ,{\bf x})
= {C\over (2\pi )^{3/2}} \int_{-\infty}^\infty d^3{\bf n}
  \sum_{s=1}^2~{\stackrel{s}{p}}_{ij} ({\bf n})
   {1\over \sqrt{2n}}
\left[ {\stackrel{s}{h}}_n (\eta ) e^{i{\bf nx}}~
                 {\stackrel{s}{c}}_{\bf n}
                +{\stackrel{s}{h}}_n^{\ast}(\eta ) e^{-i{\bf nx}}~
                 {\stackrel{s}{c}}_{\bf n}^{\dag}  \right],
\end{eqnarray}
where $C$ is a constant which will be discussed later.  
The creation and annihilation operators satisfy the conditions 
$[{\stackrel{s'}{c}}_{\bf n},~{\stackrel{s}{c}}_{{\bf m}}^{\dag}]=
\delta_{s's}\delta^3({\bf n}-{\bf m})$, 
${\stackrel{s}{c}}_{\bf n}|0\rangle =0$, where $|0\rangle$ 
(for each ${\bf n}$ and $s$) is the fixed 
initial vacuum state discussed below. 
The wave number $n$ is related with the wave vector ${\bf n}$ by 
$n = (\delta_{ij}n^in^j)^{1/2}$. 
The two polarization tensors ${\stackrel{s}{p}}_{ij}({\bf n})$ $(s = 1, 2)$ 
obey the conditions
\[
 {\stackrel{s}{p}}_{ij}n^j = 0, ~~
 {\stackrel{s}{p}}_{ij}\delta^{ij} = 0, ~~
 {\stackrel{s'}{p}}_{ij}
 {\stackrel{s}{p}}~^{ij} = 2\delta_{ss'}, ~~
 {\stackrel{s}{p}}_{ij}(-{\bf n}) = {\stackrel{s}{p}}_{ij}({\bf n}).
\]
The time evolution, one and the same for all ${\bf n}$ belonging to a 
given $n$, is represented by the complex time-dependent function
${\stackrel{s}{h}}_n(\eta )$. This evolution is dictated by the
Einstein equations. The nonlinear nature of the Einstein equations
leads to the coupling of 
${\stackrel{s}{h}}_n(\eta )$ with the pump field $a(\eta)$.
For every wave number $n$  
and each polarization component $s$,
the functions ${\stackrel{s}{h}}_n(\eta )$ have the form 
\begin{equation}
\label{7}
  {\stackrel{s}{h}}_n(\eta ) = {1\over a(\eta )} 
  [{\stackrel{s}{u}}_n(\eta ) + {\stackrel{s}{v}}_n^{\ast} (\eta )],  
\end{equation}
where ${\stackrel{s}{u}}_n(\eta )$ and ${\stackrel{s}{v}}_n(\eta )$
can be expressed in terms of the three
real functions (the polarization index $s$ is omitted): 
$r_n$ - squeeze parameter, 
$\phi_n$ - squeeze angle, $\theta_n$ - rotation angle,   
\begin{equation}
\label{8}
   u_n = e^{i{\theta}_n} \cosh~{r}_n, \qquad
   v_n = e^{-i({\theta}_n - 2{\phi}_n )} \sinh~{r}_n.
\end{equation}
The dynamical equations for $u_n(\eta)$ and $v_n(\eta)$ 
\begin{equation}
\label{9}
i\frac{{\rm d} u_n}{{\rm d}\eta} = n u_n + i\frac{a'}{a} v_n^{*}, \qquad 
i\frac{{\rm d} v_n}{{\rm d}\eta} = n v_n + i\frac{a'}{a} u_n^{*}
\end{equation}
lead to the dynamical equations governing 
the functions $r_n(\eta)$, $\phi_n(\eta)$, $\theta_n(\eta)$ \cite{g2}: 
\begin{equation}
\label{10}
r_n^{\prime} = \frac{a^{\prime}}{a} \cos{2{\phi}_n}, \quad
\phi_n^{\prime} = -n - \frac{a^{\prime}}{a} \sin{2{\phi}_n}\coth~2{r}_n, \quad 
\theta_n^{\prime} = -n - \frac{a^{\prime}}{a} \sin{2{\phi}_n}\tanh~{r}_n,
\end{equation}
where $^{\prime} = {\rm d}/{\rm d}\eta$, and the evolution begins 
from $r_n = 0$. This value of $r_n$ 
characterizes the initial vacuum state $|0\rangle$ which is defined 
long before the interaction with the pump field
became effective, that is, long before the coupling term 
$a^{\prime}/a$ became comparable with $n$. 
The constant $C$ should be taken as  
$C=\sqrt{16\pi}~l_{Pl}$ where $l_{Pl}=(G\hbar /c^3)^{1/2}$ is 
the Planck length. This particular value of the constant $C$
guarantees the correct quantum normalization of the field: energy
$\frac{1}{2} \hbar \omega$ per each mode in the initial vacuum state.  
The dynamical equations and their
solutions are identical for both polarization components $s$. 
\par
Equations (\ref{9})
can be translated into the more familiar form of the second-order  
differential equation for the function 
${\stackrel{s}{\mu}}_n(\eta ) \equiv 
{\stackrel{s}{u}}_n(\eta ) + {\stackrel{s}{v}}_n^{\ast} (\eta )
\equiv a(\eta) {\stackrel{s}{h}}_n(\eta )$ \cite{g1}:
\begin{equation}
\label{11}
\mu_{n}^{\prime\prime} + \mu_{n} \left[n^2 - 
\frac{a^{\prime\prime}}{a}\right] = 0.    
\end{equation}
\par
Clearly, this is the equation for a parametrically disturbed oscillator
(compare with Eq. (\ref{3})). In absence of the gravitational parametric
influence represented by the term  
$a^{\prime\prime}/a$, the frequency of the oscillator defined in terms
of $\eta$-time would be a constant: $n$. Whenever the term    
$a^{\prime\prime}/a$ can be neglected, the general solution 
to Eq. (\ref{11}) has the usual oscillatory form
\begin{equation}
\label{12}
\mu_n (\eta) = A_n e^{-in\eta} + B_n e^{in\eta},
\end{equation}
where the constants $A_n$, $B_n$ are determined by the initial conditions.   
On the other hand, whenever the term $a^{\prime\prime}/a$ is dominant, 
the general solution to Eq. (\ref{11}) has the form
\begin{equation}
\label{13}
\mu_n (\eta) = C_n a + D_n a \int^{\eta}\frac{{\rm d} \eta}{a^2}.
\end{equation}
In fact, this approximate solution is valid as long as $n$ is small in
comparison with $|a^{\prime}/a|$. This is more clearly seen from the
equivalent form of Eq. (\ref{11}) written in terms of the 
function $h_n (\eta)$ \cite{ll}:
\begin{equation}
\label{14}
h_n^{\prime\prime} + 2\frac{a^{\prime}}{a} h_n^{\prime} + n^2 h_n = 0.
\end{equation} 
For growing functions $a(\eta)$, that is, in expanding universes, the
second term in Eq.(\ref{13}) is usually smaller than the first one (see below), 
so that, as long as $n \ll a^{\prime}/a$, the dominant solution is the growing
function $\mu_n (\eta) = C_n a(\eta)$, and 
\begin{equation}
\label{d}
h_n = const.
\end{equation} 
\par
Equation (\ref{11}) can be also looked at as a kind of the Schrodinger 
equation for a particle moving in presence of the effective 
potential $U(\eta) = a^{\prime\prime}/a$. In the situations that
are normally considered, the potential
$U(\eta)$ has a bell-like shape and forms a barrier (see $Fig. 3$). 
When a given mode $n$ is outside the barrier, its amplitude $h_n$ 
is adiabatically decreasing with time: 
$h_n \propto \frac{e^{{\pm}in\eta}}{a(\eta)}$. 
This is shown in $Fig. 3$ by oscillating lines with decreasing amplitudes 
of oscillations. The modes with sufficiently high frequencies do not 
interact with the barrier, they stay above the barrier. Their amplitudes 
$h_n$ behave adiabatically all the time. For
these high-frequency modes, the initial vacuum state (in the Schrodinger 
picture) remains the vacuum forever. On the other hand, the modes that 
interact with the barrier are subject to the superadiabatic amplification.   
Under the barrier and as long as   
$n < a'/a$,  the function $h_n$ stays constant instead of the adiabatic
decrease. For these modes, the initial vacuum state evolves into a
squeezed vacuum state.

\begin{figure}
\epsfxsize=0.8\textwidth
\centerline{\epsfbox{barrier.fig3}} 
\end{figure}

After having formulated the initial conditions, the present day behaviour 
of $r_n$, $\phi_n$, $\theta_n$ (or, equivalently, the present day behaviour
of $h_n$) is  
essentially all we need to find. The mean number of particles
in a two-mode squeezed state is $2\sinh^2{r_n}$ for each $s$. This number 
determines the 
mean square amplitude of the gravitational wave field. The time behaviour of
the squeeze angle $\phi_n$ determines the time dependence of the correlation 
functions of the field. The amplification (that is, the growth of $r_n$) 
governed 
by Eq. (\ref{10}) is different for different wave numbers $n$. Therefore, 
the present day results depend on the present day frequency $\nu$
($\nu = {cn}/{2 \pi a}$) measured in $Hz$.  
\par 
In cosmology, the function $H \equiv \dot a/a \equiv c a^{\prime}/a^2$ 
is the time-dependent Hubble parameter. The function $l \equiv c/H$
is the time-dependent Hubble radius. The time-dependent wavelength of the 
mode $n$ is $\lambda = 2 \pi a/n$. The wavelength $\lambda$ has this universal
definition in all regimes. In contrast, the $\nu$ defined 
as $\nu = {cn}/{2 \pi a}$ has the usual meaning of a frequency of an 
oscillating process only in the short-wavelength (high-frequency) regime of
the mode $n$, that is, in the regime where $\lambda \ll l$. 
As we have seen above, the qualitative
behaviour of solutions to Eqs. (\ref{11}), (\ref{14}) depends 
crucially on the comparative  
values of $n$ and $a'/a$, or, in other words, on the comparative
values of $\lambda(\eta)$ and $l(\eta)$. This relationship is also 
crucial for solutions to Eq. (\ref{10}) as we shall see now. 
\par
In the short-wavelength regime, that is, during intervals of time
when the wavelength $\lambda(\eta)$ is shorter than the Hubble 
radius $l(\eta) = a^2/a^{\prime}$, the term $n$ in (\ref{10}) is dominant. 
The functions $\phi_n(\eta)$ and 
$\theta_n(\eta)$ are $\phi_n = -n(\eta + \eta_n)$, $\theta_n = \phi_n$
where $\eta_n$ is a constant.  
The factor $\cos 2\phi_n$ is a quickly oscillating
function of time, so the squeeze parameter $r_n$ stays practically constant.
This is the adiabatic regime for a given mode. 
\par
In the opposite, long-wavelength regime, the term $n$ can be 
neglected.
The function $\phi_n$ is $\tan \phi_n(\eta) \approx const/a^2(\eta)$,
and the squeeze angle
quickly approaches one of the two values: $\phi_n = 0$ or $\phi_n = \pi$   
(analog of ``phase bifurcation" \cite{w}). 
The squeeze parameter $r_n(\eta)$ grows with time according to   
\begin{equation}
\label{15}
r_n(\eta) \approx ln \frac{a(\eta)}{a_*}~, 
\end{equation} 
where $a_*$ is the value of $a(\eta)$ at $\eta_*$, when the 
long-wavelength regime, for a given $n$, begins. The final amount of $r_n$ is
\begin{equation}
\label{16}
r_n \approx ln \frac{a_{**}}{a_*}~, 
\end{equation} 
where $a_{**}$ is the value of $a(\eta)$ at $\eta_{**}$, when the 
long-wavelength regime and 
amplification come to the end. It is important to emphasize that it is not
a ``sudden transition" from one cosmological era to another that is responsible
for amplification, but the entire interval of the long-wavelength 
(non-adiabatic) regime.
\par
After the end of amplification, the accumulated
(and typically large) squeeze parameter $r_n$ stays approximately constant.
The mode is again in the adiabatic regime. In course of the evolution, the 
complex functions 
${\stackrel{s}{u}}_n(\eta ) + {\stackrel{s}{v}}_n^{\ast}(\eta )$
become practically real, and one has 
${\stackrel{s}{h}}_n(\eta ) \approx {\stackrel{s}{h}}_n^{\ast}(\eta ) \approx
\frac{1}{a} e^{r_n} \cos \phi_n(\eta)$.  
Every amplified mode $n$ of the field (\ref{6}) takes the form of 
a product of a function of time and a (random, operator-valued) 
function of spatial coordinates; the mode acquires a 
standing-wave pattern. The periodic dependence $\cos \phi_n(\eta)$
will be further discussed below.  
\par 
It is clearly seen from the fundamental equations (\ref{10}), (\ref{11}), 
(\ref{14}) that 
the final results depend only on $a(\eta)$. Equations do not ask us the 
names of our favorite cosmological prejudices, they ask us about the
pump field $a(\eta)$. Conversely, from the measured relic gravitational waves,
we can deduce the behaviour of $a(\eta)$, which is essentially the purpose
of detecting the relic gravitons.

\section{Cosmological Pump Field}

With the chosen initial conditions, the final 
numerical results for relic gravitational waves depend on the concrete 
behaviour of the pump
field represented by the cosmological scale factor $a(\eta)$. We know 
a great deal about $a(\eta)$. We know that $a(\eta)$ behaves as 
$a(\eta) \propto \eta^2$
at the present matter-dominated stage. We know that this stage
was preceeded by the radiation-dominated stage $a(\eta) \propto \eta$.
At these two stages of evolution the functions $a(\eta)$ are simple 
power-law functions of $\eta$.   
What we do not know is the function $a(\eta)$ describing the initial stage 
of expansion of the very early Universe, that is, before
the era of primordial nucleosynthesis. It is convenient to
parameterize $a(\eta)$ at this initial stage also by power-law functions
of $\eta$. First, this is a sufficiently broad class of functions, which, 
in addition, allows us to find exact solutions to our fundamental 
equations. Second, it is known \cite{g1} that
the pump fields $a(\eta)$ which have power-law dependence in terms of $\eta$, 
produce gravitational waves with simple power-law spectra in terms of $\nu$.
These spectra are easy to analyze and discuss in the context of detection. 
\par
We model cosmological expansion by several successive eras. Concretely,   
we take $a(\eta)$ at the initial stage of expansion ($i$-stage) as
\begin{equation}
\label{17}
a(\eta) = l_o|\eta|^{1 + \beta},  
\end{equation}
where $\eta$ grows from $- \infty$,  and $1 + \beta < 0$. We will show
later how the available observational data constrain the 
parameters $l_o$ and $\beta$. The $i$-stage lasts up to a certain   
$\eta = \eta_1$, $\eta_1 < 0$. To make our analysis more general,
we assume that the $i$-stage was followed by some interval of the 
$z$-stage ($z$ from Zeldovich). It is known that an
interval of evolution governed by the most ``stiff" matter
(effective equation of state $p = \epsilon$) advocated by Zeldovich, 
leads to a relative increase of gravitational wave amplitudes \cite{g1}.    
It is also known that the requirement of conistency of the graviton production 
with the observational restrictions does not allow the ``stiff" matter interval
to be too much long \cite{g1}, \cite{zn}. However, we want to 
investigate any interval of
cosmological evolution that can be consistently included. In fact,  
the $z$-stage of expansion that we include is quite general. 
It can be governed by a ``stiffer than radiation" \cite{gio} matter, 
as well as by a ``softer than
radiation" matter. It can also be simply a part of the radiation-dominated 
era. Concretely, we take $a(\eta)$ at the interval of time from $\eta_1$ 
to some $\eta_s$ ($z$-stage) in the form  
\begin{equation}
\label{18}
a(\eta) = l_o a_z (\eta - \eta_p)^{1 + \beta_s},
\end{equation}  
where $1 + \beta_s > 0$. For the particular choice
$\beta_s = 0$, the $z$-stage reduces to an interval
of expansion governed by the radiation-dominated matter. Starting 
from $\eta_s$ and up to $\eta_2$ the Universe was governed by the 
radiation-dominated matter ($e$-stage). So, at this interval of evolution, 
we take the scale factor in the form   
\begin{equation}
\label{19}
a(\eta) = l_oa_e(\eta - \eta_e). 
\end{equation} 
And, finally, from $\eta =\eta_2$ the expansion went over into the
matter-dominated era ($m$-stage):   
\begin{equation}
\label{20}
a(\eta) = l_oa_m(\eta - \eta_m)^2. 
\end{equation} 
A link between the arbitrary constants participating in 
Eqs. (\ref{17}) - (\ref{20}) is provided
by the conditions of continuous joining of the functions $a(\eta)$ 
and $a^{\prime}(\eta)$ at points of transitions $\eta_1$, $\eta_s$, $\eta_2$.
\par
We denote the present time by $\eta_R$ ($R$ from reception). This time is
defined by the observationally known value of the present-day Hubble
parameter $H(\eta_R)$ and Hubble radius $l_H=c/{H(\eta_R)}$. 
For numerical estimates we will be using  
$l_H  \approx 2 \times 10^{28}~{\rm cm}$. It is convenient to choose   
$\eta_R - \eta_m = 1$, so that $a(\eta_R) = 2l_H$. The ratio 
\[
a(\eta_R)/a(\eta_2) \equiv \zeta_2
\]
is believed to be around $\zeta_2 = 10^4$. We also denote   
\[
a(\eta_2)/a(\eta_s) \equiv \zeta_s~,~~~ a(\eta_s)/a(\eta_1) \equiv \zeta_1~.  
\]
With these definitions, all the constants participating in 
Eqs. (\ref{17}) - (\ref{20}) (except
parameters $\beta$ and $\beta_s$ which should be chosen from
other considerations)
are being expressed in terms of $l_H$, $\zeta_2$, $\zeta_s$,
and $\zeta_1$. For example,
\[
|\eta_1|= \frac{|1+\beta|}{2\zeta_2^{\frac{1}{2}}
\zeta_s{\zeta_1}^{\frac{1}{1+\beta_s}}} ~.
\]
The important constant $l_o$ is expressed as 
\begin{equation}
\label{21}
l_o = b l_H\zeta_2^{\frac{\beta-1}{2}}\zeta_s^{\beta}{\zeta_1}^{\frac{\beta-\beta_s}{1+\beta_s}} ,  
\end{equation}  
where $b \equiv 2^{2+\beta}/{|1+\beta|}^{1+\beta}$. Note that $b=1$ for
$\beta = -2$. 
[This expression for $l_o$ may help to relate formulas written here with 
the equivalent treatment \cite{g3} which was given in slightly 
different notations.] 
The sketch of the entire evolution $a(\eta)$ is given in $Fig. 4$.

\begin{figure}
\epsfxsize=0.8\textwidth
\centerline{\epsfbox{scalefac.fig4}} 
\end{figure}

\par
We work with the spatially-flat models (\ref{4}).
At every instant of time, the energy density $\epsilon(\eta)$ of matter 
driving the evolution is related with the Hubble radius $l(\eta)$ by
\begin{equation}
\label{22}
\kappa \epsilon(\eta) = \frac{3}{l^2(\eta)}, 
\end{equation}
where $\kappa = 8\pi G/ c^4$.
For the case of power-law scale factors $a(\eta) \propto \eta^{1+\beta}$, 
the effective pressure $p(\eta)$ of the matter is related with the
$\epsilon(\eta)$ by the effective equation of state 
\begin{equation}
\label{23}
p = \frac{1-\beta}{3(1+\beta)} \epsilon.
\end{equation}
For instance, $p=0$ for $\beta =1$, $p=\frac{1}{3} \epsilon$ for $\beta=0$,
$p= -\epsilon$ for $\beta = -2$, and so on. Each interval of
the evolution (\ref{17})-(\ref{20}) is governed by one of these 
equations of state.   
\par
In principle, the function $a(\eta)$ could be even more complicated than 
the one that we consider.  
It could even include an interval of the early contraction, 
instead of expansion, leading to the ``bounce" of the scale factor.  
In case of a decreasing $a(\eta)$ the gravitational-wave equation can 
still be analyzed and the amplification
is still effective \cite{g1}. However, the Einstein equations for
spatially-flat models do not permit a 
regular ``bounce" of $a(\eta)$ (unless $\epsilon$ vanishes at the moment
of ``bounce"). Possibly, a ``bounce" solution can be realized in alternative
theories, such, for example, as string-motivated cosmologies 
\cite{vg}. For a  
recent discussion of spectral slopes of gravitational waves produced
in ``bounce" cosmologies, see~\cite{cr}.

\section{Solving Gravitational Wave Equations}

The evolution of the scale factor $a(\eta)$ given by 
Eqs. (\ref{17}) - (\ref{20}) and
sketched in $Fig. 4$ allows us to calculate the function $a'/a$. This function
is sketched in $Fig. 5$. In all the theoretical generality, the left-hand-side
of the barrier in $Fig. 5$ could also consist of several pieces, but we
do not consider this possibility here. The graph also shows the 
important wave numbers
$n_H$, $n_2$, $n_s$, $n_1$. The $n_H$ marks the wave whose today's wavelength 
$\lambda(\eta_R)= 2 \pi a(\eta_R)/n_H$ is equal to the today's Hubble 
radius $l_H$. With our parametrization $a(\eta_R) = 2 l_H$, this wavenumber
is $n_H = 4 \pi$. The $n_2$ marks the wave whose wavelength $\lambda(\eta_2)=
2 \pi a(\eta_2)/n_2$ at $\eta= \eta_2$ is equal to the Hubble 
radius $l(\eta_2)$ at $\eta = \eta_2$. Since
$\lambda(\eta_R)/\lambda(\eta_2) = (n_2/n_H)[a(\eta_R)/a(\eta_2)]$ and 
$l(\eta_R)/l(\eta_2) = [a(\eta_R)/a(\eta_2)][a(\eta_R)/a(\eta_2)]^{1/2}$,
this gives us $n_2/n_H = [a(\eta_R)/a(\eta_2)]^{1/2} = \zeta_2^{1/2}$. 
Working out in a similar fashion other ratios, we find 
\begin{equation}       
\label{24}
\frac{n_2}{n_H} = \zeta_2^{\frac{1}{2}},~~~ \frac{n_s}{n_2} = \zeta_s,
~~~\frac{n_1}{n_s} =\zeta_1~^{\frac{1}{1+\beta_s}}. 
\end{equation}

\par
Solutions to the gravitational wave equations exist for any $a(\eta)$.
At intervals of power-law dependence $a(\eta)$, solutions to Eq. (\ref{11}) 
have simple form of the Bessel functions. We could have found piece-wise 
exact solutions to Eq. (\ref{11}) and join them in the transition points. 
However, we will use a much simpler treatment which is sufficient for our 
purposes. We know that the
squeeze parameter $r_n$ stays constant in the short-wavelength regimes and
grows according to Eq. (\ref{15}) in the long-wavelength regime. All modes 
start in the vacuum state, that is, $r_n = 0$ initially. After the end of 
amplification, the accumulated value (\ref{16}) stays constant up to today.
To find today's value of $e^{r_n}$ we need to calculate the 
ratio $a_{**}(n)/a_*(n)$. For
every given $n$, the quantity $a_*$ is determined by the condition
$\lambda(\eta_*) = l(\eta_*)$, wheras $a_{**}$ is determined by the
condition $\lambda(\eta_{**}) = l(\eta_{**})$.  

\begin{figure}
\epsfxsize=0.8\textwidth
\centerline{\epsfbox{aprime-a.fig5}} 
\end{figure}

\par
Let us start from the mode $n = n_1$. For this wave number we have 
$a_* = a_{**} = a (\eta_1)$, and therefore $r_{n_1} = 0$. The higher frequency 
modes $n > n_1$ (above the barrier in $Fig. 5$) have never been in the 
amplifying regime, so we can write
\begin{equation}
\label{25} 
e^{r_n} = 1,~~ n\ge n_1.
\end{equation}
Let us now consider the modes $n$ in the interval $n_1 \ge n \ge n_s$. 
For a given $n$ we need to know $a_* (n)$ and $a_{**}(n)$. Using Eq. (\ref{17})
one finds $a_*(n)/a_*(n_1) = (n_1/n)^{1 +\beta}$, and using Eq. (\ref{18}) one 
finds $a_{**}(n)/a_{**}(n_s) = (n_s/n)^{1 +\beta_s}$. Therefore, one finds 
\[
\frac{a_{**}(n)}{a_*(n)} = \frac{a_{**}(n_s)}{a_*(n_1)} 
\left(\frac{n_s}{n}\right)^{1+\beta_s}\left(\frac{n}{n_1}\right)^{1+\beta}.   
\]
Since $a_{**}(n_s) = a(\eta_s)$, $a_*(n_1) = a(\eta_1)$, and 
$a(\eta_s)/a(\eta_1) = \zeta_1 =(n_1/n_s)^{1+\beta_s}$, we arrive at
\[
\frac{a_{**}(n)}{a_*(n)} = \left(\frac{n}{n_1}\right)^{\beta - \beta_s}.   
\]
Repeating this analysis for other intervals of the decreasing $n$, 
we come to the conclusion that
\begin{eqnarray}
\label{26} 
e^{r_n}& =& \left(\frac{n}{n_1}\right)^{\beta - \beta_s},~~ 
n_1 \ge n \ge n_s, \nonumber \\ 
e^{r_n}& =& \left(\frac{n}{n_s}\right)^{\beta}
\left(\frac{n_s}{n_1}\right)^{\beta - \beta_s} ,~~ 
n_s \ge n \ge n_2, \nonumber \\
e^{r_n}& =& \left(\frac{n}{n_2}\right)^{\beta - 1}
\left(\frac{n_2}{n_1}\right)^{\beta}\left(\frac{n_s}{n_1}\right)^{-\beta_s},  
 n_2 \ge n \ge n_H. 
\end{eqnarray}
The mnemonic rule of constructing $e^{r_n}$ at successive intervals 
of decreasing $n$ is simple.
If the interval begins at $n_x$, one takes $(n/n_x)^{\beta_{*} - \beta_{**}}$
and multiples with $e^{r_{n_x}}$, that is, with the previous interval's 
value $e^{r_n}$ calculated at the end of that interval $n_x$. 
For the function $a'/a$ that we are working with, the $\beta_{*}$ is always
$\beta$, whereas the $\beta_{**}$ takes the values $\beta_s$, $0$, $1$ at
the successive intervals.   
\par
The modes with $n < n_H$ are still in the long-wavelength regime. For these
modes, we should take $a(\eta_R)$ instead of $a_{**}(n)$. Combining with
$a_*(n)$, we find
\begin{equation}
\label{27}
e^{r_n} = \left(\frac{n}{n_H}\right)^{\beta +1}
\left(\frac{n_H}{n_2}\right)^{\beta - 1}
\left(\frac{n_2}{n_1}\right)^{\beta}\left(\frac{n_s}{n_1}\right)^{-\beta_s},  
~~  n \le n_H.
\end{equation}
Formulas (\ref{25}) - (\ref{27}) give approximate values of $r_n$ 
for all $n$. The
factor $e^{r_n}$ is $e^{r_n} \ge 1$ for $n \le n_1$, and  
$e^{r_n} \gg 1$ for $n \ll n_1$. This factor determines the mean square
amplitude of the gravitational waves.  
\par
The mean value of the field $h_{ij}$ is zero at every moment of time $\eta$
and in every spatial point ${\bf x}$:                         
$\langle 0|h_{ij}(\eta, {\bf x})|0\rangle = 0$.
The variance 
\[
\langle 0|h_{ij}(\eta, {\bf x})h^{ij}(\eta, {\bf x})|0\rangle ~\equiv~
\langle h^2 \rangle 
\] 
is not zero, and it determines the mean square amplitude of the generated
field - the quantity of interest for the experiment. Taking the product
of two expressions (\ref{6}) one can show that   	 
\begin{equation}
\label{28}
\langle h^2 \rangle = \frac{C^2}{2\pi^2} \int_0^\infty n \sum_{s=1}^2
\Big| {\stackrel{s}{h}}_n(\eta )\Big|^2 ~{\rm d}n \equiv 
\int_0^\infty h^2(n, \eta) \frac{{\rm d}n}{n}~. 
\end{equation}
Using the representation (\ref{7}), (\ref{8}) in Eq. (\ref{28}) one can 
also write
\begin{equation}
\label{29}
\langle h^2 \rangle = 
\frac{C^2}{{\pi^2}a^2 } \int_0^\infty n {\rm d}n (\cosh2{r_n} + 
\cos2{\phi_n}\sinh2{r_n}). 
\end{equation}
We can now consider the present era and use the fact that $e^{r_n}$ are 
large numbers for all $n$ in the interval of our interest 
$n_1 \ge n \ge n_H$. Then, we can derive  
\begin{equation}
\label{30}
h(n, \eta)\approx \frac{C}{\pi}\frac{1}{a(\eta_R)}n e^{r_n}\cos\phi_n(\eta)=
8 \sqrt{\pi}\left(\frac{l_{Pl}}{l_H}\right)\left(\frac{n}{n_H}\right)
e^{r_n}\cos\phi_n(\eta) ~.
\end{equation}
The quantity $h(n, \eta)$ is the
dimensionless spectral amplitude of the field whose numerical value
is determined by the calculated squeeze parameter $r_n$.  
The oscillatory factor $\cos\phi_n(\eta)$ reflects the squeezing (standing
wave pattern) acquired by modes with $n_1 > n > n_H$. For modes with
$n < n_H$ this factor is approximately $1$. For high-frequency modes 
$n \gg n_H$ one has
$\phi_n(\eta) \approx n(\eta - \eta_n) \gg 1$, so that $h(n, \eta)$ makes 
many oscillations while the scale factor $a(\eta)$ is practically fixed 
at $a(\eta_R)$.
\par 
The integral (\ref{29}) extends formally from $0$ to $\infty$. 
Since $r_n \approx 0$
for $n \ge n_1$, the integral diverges at the upper limit. This is a typical
ultra-violet divergence. It should be discarded (renormalized to zero) 
because it comes from 
the modes which have always been in their vacuum state. At the lower limit,
the integral diverges, if $\beta \le -2$. This is an infra-red divergence
which comes from the assumption that the amplification process has started
from infinitely remote time in the past. One can deal with this 
divergence either by introducing a lower frequency cut-off (equivalent to
the finite duration of the amplification) or by considering only the
parameters $\beta > -2$, in which case the integral is convergent at the lower
limit.
It appears that the available observational data (see below) favour this
second option. The particular case $\beta = -2$ corresponds to  
the de Sitter evolution $a(\eta) \propto |\eta|^{-1}$. In this case, 
the $h(n)$ found in Eqs. (\ref{30}), (\ref{27}) does not depend on $n$. 
This is known as the Harrison-Zeldvich, or scale-invariant, spectrum. 
\par
An alternative derivation of the spectral amplitude $h(n)$ uses the
approximate solutions (\ref{12}), (\ref{13}) to the wave equation (\ref{11}). 
This method gives exactly the same, as in Eqs. (\ref{30}),
(\ref{25}) - (\ref{27}) numerical values of $h(n)$, but does not 
reproduce the oscillatory factor $\cos\phi_n(\eta)$. 
\par
One begins with the initial spectral amplitude $h_i(n)$ defined 
by quantum normalization:
$h_i(n) = 8 \sqrt{\pi} (l_{Pl}/ \lambda_i)$. This is the amplitude of the mode
$n$ at the moment $\eta_*$ of entering the long wavelength regime, i.e.  
when the mode's wavelength $\lambda_i$ is equal to the Hubble 
radius $l(\eta_*)$. For $\lambda_i$ one derives 
\begin{equation}
\label{31}
\lambda_i = \frac{1}{b} l_o \left(\frac{n_H}{n}\right)^{2+\beta}. 
\end{equation}
Thus, we have
\begin{equation}
\label{32}
h_i(n) = A \left(\frac{n}{n_H}\right)^{2+\beta},
\end{equation}
where $A$ denotes the constant  
\begin{equation}
\label{33}
A = b 8\sqrt{\pi} \frac{l_Pl}{l_o}~. 
\end{equation} 
The numbers $h_i(n)$ are defined at the beginning of the long-wavelength
regime. In other words, they are given along the left-hand-side slope of the 
barrier in $Fig. 5$.
We want to know the final numbers (spectral amplitudes) $h(n)$ 
which describe the field today, at $\eta_R$. 
\par
According to the dominant solution $h_n(\eta) =const$ of the long-wavelength
regime (see Eq. (\ref{d})), the initial amplitude $h_i(n)$ stays practically 
constant up to the end of the long-wavelength regime at $\eta_{**}$, 
that is, up to the right-hand-side slope of the barrier. 
[The second term in Eq. (\ref{13}) could be
important only at the $z$-stage and only for parameters 
$\beta_s \le -(1/2)$, which
correspond to the effective equations of state $p \ge \epsilon$. 
In order to keep the analysis simple, we do not consider those cases.] 
After the completion of the long-wavelength regime, the amplitudes decrease 
adiabatically in proportion to $1/a(\eta)$, up to the present time. 
Thus, we have 
\begin{equation}
\label{34}  
h(n) = A \left(\frac{n}{n_H}\right)^{2+\beta} \frac{a_{**}(n)}{a(\eta_R)}. 
\end{equation}
\par
Let us start from the lower end of the spectrum, $n \le n_H$, and go 
upward in $n$. The modes $n \le n_H$ have not started yet the adiabatic
decrease of the amplitude, so we have
\begin{equation}
\label{35}  
h(n) = A \left(\frac{n}{n_H}\right)^{2+\beta},~~ n \le n_H. 
\end{equation}
Now consider the interval $n_2 \ge n \ge n_H$. At this interval,
the $a_{**}(n)/a(\eta_R)$ scales as $(n_H/n)^2$, so we have
\begin{equation}
\label{36}  
h(n) = A \left(\frac{n}{n_H}\right)^{\beta},~~ n_2 \ge n \ge n_H. 
\end{equation}
At the interval $n_s \ge n \ge n_2$ the ratio  
$a_{**}(n)/a(\eta_R) = [a_{**}(n)/a(\eta_2)][a(\eta_2)/a(\eta_R)]$ 
scales as $(n_2/n)(n_H/n_2)^2$, so we have
\begin{equation}
\label{37}  
h(n) =A\left(\frac{n}{n_H}\right)^{1+\beta}\frac{n_H}{n_2},~~ n_s\ge n\ge n_2.  
\end{equation}
Repeating the same analysis for the interval $n_1 \ge n \ge n_s$ we find
\begin{equation}
\label{38}  
h(n) =A\left(\frac{n}{n_H}\right)^{1+\beta -\beta_s}
\left(\frac{n_s}{n_H}\right)^{\beta_s}\frac{n_H}{n_2},~~ n_1\ge n\ge n_s.  
\end{equation}
It is seen from
Eq. (\ref{38}) that an interval of the $z$-stage with $\beta_s <0$ 
(the already imposed
restrictions require also $(-1/2) < \beta_s$) bends
the spectrum $h(n)$ upwards, as compared with Eq. (\ref{37}), for larger $n$.  
If one recalls the relationship (\ref{21}) between $l_o$ and $l_H$
and uses (\ref{26}), (\ref{27}) in Eq. (\ref{30}) one 
arrives exactly at Eqs. (\ref{35})-(\ref{38}) up to
the oscillating factor $\cos\phi_n(\eta)$. 
\par 
Different parts of the barrier in $Fig. 5$ are responsible for amplitudes and
spectral slopes at different intervals of $n$. The sketch of the generated 
spectrum $h(n)$ in conjunction with the form of the barrier 
is shown in $Fig. 6$.

\begin{figure}
\epsfxsize=0.8\textwidth
\centerline{\epsfbox{charam.fig6}} 
\end{figure}

The present day frequency of the oscillating modes, measured in $Hz$,
is defined as $\nu = cn/2\pi a(\eta_R)$. The lowest frequency (Hubble
frequency) is $\nu_H = c/l_H$. For numerical estimates we will be using  
$\nu_H \approx 10^{-18} Hz$. The ratios of $n$ are equal to the ratios of
$\nu$, so that, for example, $n/n_H = \nu/\nu_H$. For high-frequency
modes we will now often use the ratios of $\nu$ instead of ratios of $n$. 
\par
In addition to the spectral amplitudes $h(n)$ the generated field can be
also characterized by the spectral energy density parameter $\Omega_{g}(n)$. 
The energy density $\epsilon_g$ of the gravitational wave field is 
\[
\kappa \epsilon_g = \frac{1}{4} h^{ij}_{~,0} h_{ij,0} =
\frac{1}{4a^2} {h^{ij}}^{\prime} {h_{ij}}^{\prime}.
\]
The mean value $\langle 0|\epsilon_g (\eta, {\bf x})|0\rangle$ is given by
\begin{equation}
\label{39}
\kappa \langle \epsilon_g \rangle = \frac{1}{4a^2} 
\frac{C^2}{2\pi^2} \int_0^\infty n \sum_{s=1}^2
\Big| {{\stackrel{s}{h}}}^{\prime}_n(\eta )\Big|^2 ~{\rm d}n.
\end{equation}
For high-frequency modes, it is only the factor $e^{\pm i n\eta}$ that needs
to be differentiated by $\eta$. After avaraging out the oscillating factors, 
one gets  
$\Big| {{\stackrel{s}{h}}}^{\prime}_n\Big|^2 = 
n^2 \Big| {\stackrel{s}{h}}_n\Big|^2$, so that 
\begin{equation}
\label{40}
\kappa \langle \epsilon_g \rangle = \frac{1}{4a^2} 
\int_0^\infty n^2 h^2(n) \frac{{\rm d}n}{n}.  
\end{equation}
In fact, the high-frequency approximation, that has been used, permits 
integration over lower $n$ only up to $n_H$. And the upper limit, as was 
discussed above, is in practice $n_1$, not infinity.    
The parameter $\Omega_g$ is defined as $\Omega_g = 
\langle \epsilon_g \rangle /\epsilon$, where $\epsilon$ is given by
Eq. (\ref{22}) (critical density). So, we derive
\[
\Omega_g =  \int_{n_H}^{n_1} \Omega_g(n) \frac{{\rm d}n}{n} = 
\int_{\nu_H}^{\nu_1} \Omega_g(\nu) \frac{{\rm d}\nu}{\nu} 
\]
and
\begin{equation}
\label{41}
\Omega_g(\nu) = \frac{\pi^2}{3}h^2(\nu)\left(\frac{\nu}{\nu_H}\right)^2.
\end{equation}
\par
The dimensionless quantity $\Omega_g(\nu)$ is useful because it allows us
to quickly evaluate the cosmological importance of the generated field in
a given frequency interval. However, the primary and more universal concept
is $h(\nu)$, not $\Omega_g(\nu)$. It is the field, not its energy density,
that is directly measured by the gravity-wave detector. One should also note
that some authors use quite a misleading definition $\Omega_g (f) =
(1/\rho_c)(d \rho_{gw}/d \ln f)$ which suggests differentiation of the 
gravity-wave energy density by frequency. This would be incorrect and could
cause disagreements in numerical values of $\Omega_g$. Whenever we use
$\Omega_g(\nu)$, we mean relationship (\ref{41}); and for order 
of magnitude estimates one can use \cite{g1}:
\begin{equation}
\label{42}
\Omega_g(\nu) \approx h^2(\nu) \left(\frac{\nu}{\nu_H}\right)^2 .
\end{equation}

\section{Theoretical and Observational Constraints}

The entire theoretical approach is based on the assumption that a weak 
quantized gravity-wave field interacts with a classical pump field. We 
should follow the validity of this approximation throughout the analysis. 
The pump field can be treated as a classical gravitational field 
as long as the driving energy
density $\epsilon$ is smaller than the Planck energy density, or, in
other words, as long as the Hubble radius $l(\eta)$ is greater than the
Planck length $l_{Pl}$. This is a restriction on the pump field, but
it can be used as a restriction on the 
wavelength $\lambda_i$ of the gravity-wave mode $n$ at the 
time of entry the long-wavelength regime. If $l(\eta_*) > l_{Pl}$, then
$\lambda_i > l_{Pl}$. The $\lambda_i$ is given by Eq. (\ref{31}). 
So, we need to ensure that  
\[
b \frac{l_{Pl}}{l_o} \left(\frac{\nu}{\nu_H}\right)^{2+\beta} < 1.
\]
At the lowest-frequency end $\nu = \nu_H$ this inequality
gives $b(l_{Pl}/l_o) < 1$. In fact, the observational constraints 
(see below) give a stronger restiction: 
\begin{equation}
\label{43}
b \frac{l_{Pl}}{l_o} \approx 10^{-6} , 
\end{equation} 
which we accept. Then, at the
highest-frequency end $\nu = \nu_1$ we need to satisfy
\begin{equation}
\label{44}
\left(\frac{\nu_1}{\nu_H}\right)^{2+\beta} < 10^6.
\end{equation}
\par
Let us now turn to the generated spectral amplitudes $h(\nu)$. According
to Eq. (\ref{35}) we have $h(\nu_H) \approx b 8 \sqrt{\pi} (l_{Pl}/l_o)$. The
measured microwave beckgound anisotropies, which we discuss below, 
require this number
to be at the level of $10^{-5}$, which gives the already mentioned 
Eq. (\ref{43}). The quantity $h(\nu_1)$ at the highest 
frequency $\nu_1$ is given by Eq. (\ref{38}):
\[
h(\nu_1) = b 8\sqrt{\pi} \frac{l_{Pl}}{l_o}
\left(\frac{\nu_1}{\nu_H}\right)^{1+\beta -\beta_s}
\left(\frac{\nu_s}{\nu_H}\right)^{\beta_s}\frac{\nu_H}{\nu_2}. 
\]
Using Eq. (\ref{21}) this expression for $h(\nu_1)$ can be rewritten as 
\begin{equation}
\label{45}
h(\nu_1) =8\sqrt{\pi} \frac{l_{Pl}}{l_H}\frac{\nu_1}{\nu_H}= 
8\sqrt{\pi} \frac{l_{Pl}}{\lambda_1}, 
\end{equation}
where $\lambda_1 = c/\nu_1$. This last expression for $h(\nu_1)$ is not 
surprising: the modes with $\nu \ge \nu_1$ are still in the vacuum state,
so the numerical value of $h(\nu_1)$ is determined by quantum normalization.
\par
All the amplified modes have started with small initial amplitudes 
$h_i$, at the level of
zero-point quantum fluctuations. These amplitudes are also small today, since
the $h_i$ could only stay constant or decrease. However, even these relatively
small amplitudes should obey observational constraints. We do not want the
$\Omega_g$ in the high-frequency modes, which might affect the rate of
the primordial nucleosynthesis, to exceed the level of $10^{-5}$. This
means that $\Omega_g(\nu_1)$ cannot exceed the level of $10^{-6}$ or so.   
The use of Eq. (\ref{41}) in combination with 
$\Omega_g(\nu_1) \approx 10^{-6}$ and 
$h(\nu_1)$ from Eq. (\ref{45}), gives us the highest allowed frequency
$\nu_1 \approx 3 \times 10^{10} Hz$. We will use this value of $\nu_1$
in our numerical estimates.     
Returning with this value of $\nu_1$ to Eq. (\ref{44}) we find that parameter
$\beta$ can only be $\beta \le - 1.8$. We will be treating $\beta = -1.8$
as the upper limit for the allowed values of $\beta$. 
\par
We can now check whether the accepted parameters leave room for the 
postulated $z$-stage with $\beta_s < 0$. Using Eq. (\ref{21}) we can rewrite
Eq. (\ref{43}) in the form
\begin{equation}
\label{46}
10^{-6} \frac{l_H}{l_{Pl}} =  
\left(\frac{\nu_1}{\nu_H}\right)^{- \beta}
\left(\frac{\nu_1}{\nu_s}\right)^{\beta_s}\frac{\nu_2}{\nu_H}. 
\end{equation}
We know that $\nu_2/ \nu_H = 10^2$ and $\nu_1/\nu_s$ is not smaller 
than $1$. Substituting all the numbers in Eq. (\ref{46}) one can find that
this equation cannot be satisfied for the largest possible $\beta = -1.8$. 
In the case $\beta = -1.9$, 
Eq. (\ref{46}) is only marginally satisfied, in the sense that a significant
deviation from $\beta_s =0$ toward negative $\beta_s$ can only last for 
a relatively short time. For
instance, one can accomodate $\beta_s = -0.4$ and $\nu_s = 10^8 Hz$. 
On the other hand, if one takes  
$\beta = -2$, a somewhat longer interval of the $z$-stage with $\beta_s < 0$ 
can be 
included. For instance, Eq. (\ref{46}) is satisfied if one accepts $\nu_s =
10^{-4} Hz$ and $\beta_s = - 0.3$. This allows us to slightly increase $h(\nu)$
in the interval $\nu_s < \nu < \nu_1$, as compared with the values of
$h(\nu)$ reached in the more traditional
case $\beta = -2$, $\beta_s = 0$. In what follows, we will consider 
consequences of this assumption for the prospects of detection of the
produced gravitational wave signal.    
Finally, let us see what the available information on the microwave background 
anisotropies \cite{s}, \cite{b} allows us to conclude about the 
parameters $\beta$ and $l_o$. 
\par 
Usually, cosmologists operate with the spectral index ${\rm n}$ (not to be
confused with the wave number $n$) of primordial cosmological perturbations.
Taking into account the way in which the spectral index ${\rm n}$ is defined, 
one can relate ${\rm n}$ with the spectral index $\beta + 2$ that shows up
in Eq. (\ref{35}). The relationship between them is ${\rm n} = 2\beta+5$. 
This relationship is valid independently of the nature of cosmological
perturbations. In particular, it is valid for density perturbations,
in which case the $h(n)$ of Eq. (\ref{35}) is the dimensionless 
spectral amplitude
of metric perturbations associated with density perturbations. If primordial
gravitational waves and density perturbations were generated by the 
mechanism that 
we discuss here (the assumption that is likely to be true) than the parameter
$\beta$ that participates in the spectral index is the same one that 
participates in the scale factor of Eq. (\ref{17}). 
Primordial gravitational waves and primordial density perturbations 
with the same spectral index produce approximately the same 
lower-order multipole distributions of large-scale anisotropies.       
\par
The evaluation of the spectral 
index ${\rm n}$ of primordial perturbations have resulted 
in ${\rm n} = 1.2 \pm 0.3$ \cite{b} or even in a somewhat higher 
value. A recent analysis \cite{melch} of all
available data favors ${\rm n} = 1.2$ and the quadrupole contribution of
gravitational waves twice as large as that of density perturbations. 
One can interpret these evaluations 
as indication that the true value of ${\rm n}$ lies somewhere near 
${\rm n} = 1.2$ (hopefully, the planned new observational missions will 
determine this index more accurately). This gives us the parameter $\beta$ 
somewhere near $\beta = -1.9$. We will be using $\beta = -1.9$ in our estimates 
below, as the observationally preferred value. The parameter $\beta$ can
be somewhat larger than $\beta = -1.9$. However, as we already 
discussed, the value $\beta = -1.8$ (${\rm n} = 1.4$)      
is the largest one for which the entire approch is well posed.
The Harrison-Zeldovich spectral index ${\rm n} = 1$ corresponds to $\beta = -2$. 
\par
The observed quadrupole anisotropy of the microwave background radiation is
at the level $\delta T/ T \approx 10^{-5}$.   
The quadrupole anisotropy that would be produced by the spectrum 
(\ref{35}) - (\ref{38}) is
mainly accounted for by the wave numbers near $n_H$. Thus, 
the numerical value of the quadrupole anisotropy produced by relic 
gravitational waves is approximately equal to $A$. According to general 
physical considerations and detailed calculations \cite{gden}, the 
metric amplitudes of long-wavelength gravitational waves and density
perturbations generated by the discussed amplification mechanism are of the 
same order of magnitude. Therefore, they contribute roughly equally to the 
anisotropy at lower multipoles. This gives us the estimate $A\approx 10^{-5}$, 
that we have already used in Eq. (\ref{43}). It is not yet proven 
observationally that a significant part of the observed anosotropies
at lower multipoles is indeed provided by 
relic gravitational waves, but we can at least assume this with some degree of
confidence. It is likely that the future measurements of the microwave
background radiation will help us to verify this theoretical conclusion.   
\par
Combining all the evaluated parameters together, we show in $Fig. 7$ the 
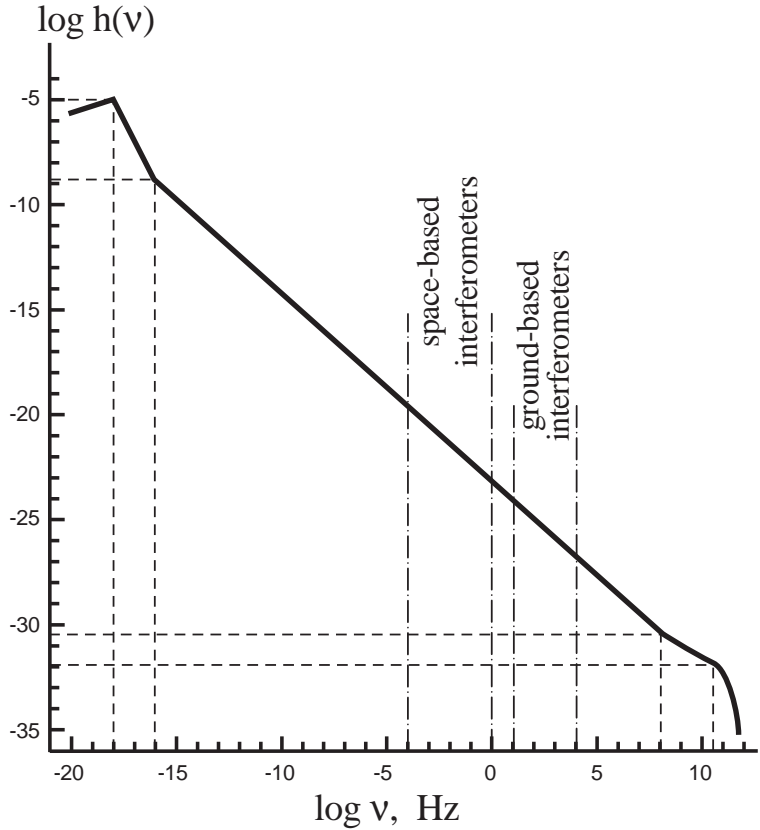
expected spectrum of $h(\nu)$ for the case $\beta = -1.9$. A small allowed
interval of the $z$-stage is also included. The intervals of the spectrum
accessible to space-based and ground-based interferometers are indicated
by vertical lines.

\begin{figure}
\epsfxsize=0.8\textwidth
\centerline{\epsfbox{predspec.fig7}} 
\end{figure}

\par
It is necessary to note \cite{gden}, \cite{g3} 
that the confirmation of any ${\rm n} > 1$ ($\beta > -2$) would 
mean that the
very early Universe was not driven by a scalar field - the cornerstone of
inflationary considerations. This is because the ${\rm n} > 1$ ($\beta > -2$) 
requires the effective equation of state at the initial stage of expansion 
to be $\epsilon + p < 0$ (see Eq. (\ref{23})), but this cannot be accomodated 
by any scalar field 
with whichever scalar field potential. The available data do not
prove yet that ${\rm n} > 1$, but this possibility seems likely. 
\par
It is also
necessary to say that a certain damage to gravitational wave research 
was inflicted by the so called ``standard inflationary result". The   
``standard inflationary
result" predicts infinitely large amplitudes of density perturbations in 
the interval of spectrum with the Harrison-Zeldovich 
slope ${\rm n} =1$ ($\beta = -2$):
$\delta\rho/\rho \propto 1/\sqrt{1 -{\rm n}}$. The metric (gravitational
field) amplitudes of density perturbations are
also predicted to be infinitely large, in the same proportion. 
Through the so-called  
``consistency relation" this divergence leads to the vanishingly small 
amplitudes of relic gravitational waves. Thus, the ``standard" inflationary 
theory predicts zero for relic gravitational waves; the spectrum similar in
shape to the one shown in $Fig. 7$ would have been shifted down by many 
orders of magnitude. 
This prediction is hanging on the ``standard inflationary result", but
the ``result" itself is in a severe conflict not only with theory but 
with observations too: when the
observers marginalize their data to ${\rm n} = 1$ (enforce this value
of ${\rm n}$ in data analysis) they find finite and small density
perturbations instead of infinitely large perturbations predicted by
inflationary theorists. 
[For analytical expressions of the ``standard inflationary result" see any
inflationary article, including recent reviews. For graphical illustration
of the predicted divergent density perturbations and quadrupole anisotropies
see, for example, \cite{ms}. For critical analysis and disagreement with 
the ``standard inflationary result" see \cite{gden}.] General relativity and
quantum field theory do not produce the ``standard inflationary result",
so we shall better return to what they say.

\section{Detectability of Relic Gravitational Waves}

We switch now from cosmology to prospects of detecting the predicted relic
gravitational waves. The ground-based \cite{Abr}-\cite{HoughD} 
and space-based \cite{lisa}, \cite{Hel} laser interferometers
(see also \cite{gw0}-\cite{gw2}) 
will be in the focus of our attention. We use laboratory frequencies 
$\nu$ and intervals of laboratory time $t$ 
$(c{\rm d}t = a(\eta_R){\rm d}\eta)$. Formulas (\ref{37}) and (\ref{38}), 
with $A = 10^{-5}$, $\nu_2/\nu_H = 10^2$, and the oscillating factor 
restored, can be written as 
\begin{equation}
\label{47}
h(\nu,t) \approx 
10^{-7}\cos[2\pi\nu(t - t_{\nu})]\left(\frac{\nu}{\nu_H}\right)^{\beta+1},
~~\nu_2 \le \nu \le \nu_s 
\end{equation} 
and
\begin{equation}
\label{48}
h(\nu,t) \approx 
10^{-7}\cos[2\pi\nu(t - t_{\nu})]
\left(\frac{\nu}{\nu_H}\right)^{1+\beta -\beta_s}
\left(\frac{\nu_s}{\nu_H}\right)^{\beta_s}.~~\nu_s \le \nu \le \nu_1 
\end{equation} 
where the deterministic (not random) constant $t_{\nu}$  
does not vary significantly from one frequency to another at 
the intervals $\Delta\nu \approx \nu$. 
The explicit time dependence of the spectral variance $h^{2}(\nu,t)$ 
of the field, or, in other words, the explicit time dependence of 
the (zero-lag) temporal correlation function of the field at every
given frequency, demonstrates 
that we are dealing with a non-stationary process (a consequence
of squeezing and severe reduction of the phase uncertainty). 
We will first ignore the oscillating factor and will compare the predicted
amplitudes with the sensitivity curves of advanced detectors. The potential
reserve of improving the signal to noise ratio by expoloiting the squeezing
will be discussed later.
\par
Let us start from the Laser Interferometer Space Antenna (LISA) \cite{lisa}.
The instrument will be most sensitive in the interval, roughly, from
$10^{-3} Hz$ to $10^{-1} Hz$, and will be reasonably sensitive
in a broader range, up to frequencies
$10^{-4} Hz$ and $1 Hz$. The sensitivity graph of LISA to a stochastic
background is usually plotted under the assumption of a 1-year observation
time, that is, the root-mean-square (r.m.s.) instrumental noise is 
being evaluated in frequency 
bins $\Delta \nu = 3\times 10^{-8} Hz$ around each frequency $\nu$.  
We need to rescale our predicted amplitude $h(\nu)$ to these bins. 
\par
The mean square amplitude of the gravitational wave field is given by
the integral (\ref{28}). Thus, the r.m.s. amplitude in the band $\Delta \nu$
centered at a given frequency $\nu$ is given by the expression
\begin{equation}
\label{49}
h(\nu, \Delta\nu) = h(\nu) \sqrt{\frac{\Delta\nu}{\nu}}.
\end{equation}
We use Eqs. (\ref{47}), (\ref{48}) and  calculate expression (\ref{49}) 
assuming $\Delta \nu = 3\times 10^{-8} Hz$. The results are plotted 
in $Fig. 8$. 
Formula (\ref{47}) has been used 
throughout the covered frequency interval for the realistic
case $\beta = -1.9$ and for the extreme case $\beta = -1.8$. 
The line marked $z$-model describes the signal produced in the 
composite model with $\beta = -2$ up to $\nu_s = 10^{-4} Hz$ 
(formula (\ref{47})) and then followed by formula (\ref{48}) 
with $\beta_s = -0.3$. This
model gives the signal a factor of $3$ higher at $\nu= 10^{-3} Hz$, 
than the model $\beta = -2$ extrapolated down to this frequency.    
\begin{figure}
\epsfxsize=0.8\textwidth
\centerline{\epsfbox{lisa.fig8}} 
\end{figure}
\par
There is no doubt that the signal $\beta = -1.8$ would be easily
detectable even with a single instrument. The signal $\beta = -1.9$
is marginally detectable, with the signal to noise ratio around $3$ or
so, in a quite narrow frequency interval near and above the frequency 
$3\times 10^{-3} Hz$. However, at lower frequencies one would
need to be concerned with the possible gravitational wave noise from 
unresolved binary stars in our Galaxy. The further improvement of the
expected LISA sensitivity by a factor of $3$ may prove to be crucial
for a confident detection of the predicted signal with $\beta = -1.9$.   
\par
Let us now turn to the ground-based interferometers operating
in the interval from $10 Hz$ to $10^{4} Hz$. The best sensitivity is
reached in the band around $\nu = 10^2 Hz$. We take this frequency 
as the representative frequency for comparison with the predicted signal. 
We will work directly in terms of the dimensionless quantity $h(\nu)$.
If necessary, the r.m.s. amplitude per $Hz^{1/2}$ at a given $\nu$ can be 
found simply 
as $h(\nu)/{\sqrt\nu}$. The instrumental noise will also be quoted in
terms of the dimensionless quantity $h_{ex}(\nu)$. 
\par
The expected sensitivity of the initial instruments at $\nu = 10^2 Hz$ is
$h_{ex} = 10^{-21}$ or better. The theoretical prediction at this
frequency, following from (\ref{47}), (\ref{48}) with $\beta_s = 0$, is 
$h_{th} = 10^{-23}$ for $\beta = -1.8$, 
and $h_{th} = 10^{-25}$ for $\beta = -1.9$. Therefore, the gap between
the signal and noise levels is from 2 to 4 orders of magnitude. The expected
sensitivity of the advanced interferometers, such as 
LIGO-II \cite{ligoII}, can be as high
as $h_{ex} = 10^{-23}$. In this case, the gap vanishes for the $\beta= -1.8$
signal and reduces to 2 orders of magnitude for the $\beta= -1.9$ signal.
$Fig. 9$ illustrates the expected signal in comparison with the LIGO-II 
sensitivity. Since the signal lines are plotted in terms of $h(\nu)$, the LISA
sensitivity curve (shown for periodic sources) should be raised and
adjusted in accordance with $Fig. 8$.

\begin{figure}
\epsfxsize=0.8\textwidth
\centerline{\epsfbox{ligo.fig9}} 
\end{figure}

\par
A signal below noise can be detected if the outputs
of two or more detectors can be cross correlated. [For the early esimates 
of detectability of relic gravitational waves see \cite{g4}.] The cross
correlation will be possible for ground-based 
interferometers, several of which are currently under construction. 
The gap between the signal and the noise levels 
should be covered by a sufficiently long observation time $\tau$.
The duration $\tau$ depends on whether the signal has any temporal 
signature known in advance, or not. We start from the assumption that no
temporal signatures are known in advance. In other words, we first ignore 
the squeezed nature of the relic background and work under the assumption
that the squeezing cannot be exploited to our advantage.  
\par
The response of an instrument to the incoming radiation is 	
$s(t) = F_{ij}h^{ij}$ where $F_{ij}$ depends on the position and 
orientation of the instrument. Since the $h^{ij}$ is a quantum-mechanical
operator (see Eq. (\ref{6})) we need to calculate the mean value of a quadratic
quantity. The mean value of the cross correlation of responses from two
instruments $\langle 0|s_1(t)s_2(t)|0\rangle$ will involve the overlap 
reduction function \cite{Mich}-\cite{Allen}, 
which we assume to be not much smaller than $1$ \cite{Flan}. 
The signal to noise ratio $S/N$ in the measurement of the amplitude 
of a signal with no specific known features increases as
$(\tau \nu)^{1/4}$, where $\nu$ is some characteristic central frequency. 
If the signal has features known in advance and exploited by the 
matched filtering technique, the $S/N$ increases as
$(\tau \nu)^{1/2}$.  
\par
We apply the guaranteed law $(\tau \nu)^{1/4}$ to initial and advanced
instruments at the representative frequency $\nu = 10^2 Hz$. This law  
requires a reasonably short time $\tau = 10^6~{\rm sec}$ in order
to improve the $S/N$ in initial instruments by two orders of magnitude  
and to reach the level of 
the signal with extreme spectral index $\beta = -1.8$.
The longer integration time or a better sensitivity will make the 
$S/N$ larger than 1. In the case of a realistic spectral index  
$\beta = -1.9$ the remaining gap of 4 orders of magnitude can be
covered by the combination of a significantly better sensitivity and
a longer observation time (not necessarily in one non-interrupted run).   
The sensitivity of the advanced laser interferometers, such as LIGO II, 
at the level $h_{ex} = 10^{-23}$ and the same observation time  
$\tau = 10^6~{\rm sec}$ would be sufficient for reaching the level of the
predicted signal with $\beta = -1.9$.
\par
An additional increase of $S/N$ can be achieved if the statistical
properties of the signal can be properly exploited.
Squeezing is automatically present at all frequencies from $\nu_H$ to
$\nu_1$. The squeeze parameter $r$ is larger in gravitational waves
of cosmological scales, and possibly the periodic structure in 
Eq. (\ref{30}) can be better revealed at those scales. However, we 
are interested here  
in frequencies accessible to ground based interferometers,
say, in the interval $30 Hz - 100 Hz$. If our intention were to monitor
one given frequency $\nu$ from the beginning of its oscillating regime
and up till now, then, in order to avoid the destructive interference from 
neighbouring modes during all that time,
the frequency resolution of the instrument should have been 
increadibly narrow, of the order of $10^{-18} Hz$.
Certainly, this is not something what we can, or
intend to do. Although the amplitudes of the waves have adiabatically
decreased and their frequencies redshifted since the beginning of their
oscillating regime, the general statistical properties of the discussed
signal are essentially the same now as they were $10$ years after the 
Big Bang or will be $1$ million years from now. 
\par 
The periodic structure (\ref{47}) may survive at some level in 
the instrumental window of sensitivity from $\nu_{min}$ (minimal frequency) 
to $\nu_{max}$ (maximal frequency). The mean square value of the field in 
this window is 
\begin{equation}
\label{50}
\int_{\nu_{min}}^{\nu_{max}}h^2(\nu,t)\frac{{\rm d}\nu}{\nu} =
10^{-14}\frac{1}{{\nu_H}^{2\beta+2}}\int_{\nu_{min}}^{\nu_{max}}
\cos^2[2\pi\nu(t - t_{\nu})]\nu^{2\beta+1}{\rm d}\nu~.
\end{equation}
Because of the strong dependence of the integrand on frequency, 
$\nu^{-2.6}$ or $\nu^{-2.8}$, the value of the integral (\ref{50}) 
is determined by 
its lower limit. Apparently, the search through the data should be based on 
the periodic structure that may survive at $\nu = \nu_{min}$. As an illustration,
one can consider such a narrow interval $\Delta \nu = \nu_{max} - \nu_{min}$ 
that the integral (\ref{50}) can be approximated by the formula 
\[
\int_{\nu_{min}}^{\nu_{max}}h^2(\nu,t)\frac{{\rm d}\nu}{\nu} \approx
10^{-14}\left(\frac{\nu_{min}}{{\nu_H}}\right)^{2\beta+2}
\left(\frac{\Delta \nu}{\nu_{min}}\right)\cos^2[2\pi\nu_{min}(t - t_{min})]~.
\]
Clearly, the correlation function is strictly periodic and its structure
is known in advance, in contrast to other possible signals. This is a typical
example of using the appriori information. Ideally, the gain in $S/N$ can
grow as $(\tau \nu_{min})^{1/2}$. This would significantly reduce the 
required observation time $\tau$. For a larger $\Delta \nu$, even an 
intermediate gain between the guaranteed law $(\tau \nu)^{1/4}$ and 
the law $(\tau \nu)^{1/2}$, adequate for the matched filtering 
technique, would help. This could potentially make the signal with
$\beta =-1.9$ measurable even by the initial laser interferometers.  
A straightforward application of (\ref{50}) for exploiting the squeezing 
may not be possible, as argued in the recent study \cite{AFP}, but more
sophisticated methods are not excluded. 
\par
For frequency
intervals covered by bar detectors and electromagnetic detectors, the
expected results follow from the same formulas (\ref{47}), (\ref{48}) 
and have been briefly
discussed elsewhere \cite{g4}, \cite{g3}.

\section{Conclusion} 

It would be strange, if the predicted signal at the level corresponding
to $\beta =-1.9$ were not seen by the instruments capable of its detection.
There is not so many cosmological assumptions involved in the derivation, 
that could prove wrong, thus invalidating our predictions. On the other hand, 
it would be even more 
strange (and even more interesting) if the relic gravitational waves
were detected at the level above the $\beta = -1.8$ line. This 
would mean that there is something fundamentally wrong in our 
basic cosmological premises. To summarise, it is quite possible that  
the detection of relic (squeezed) gravitational waves may be awaiting 
only the first generation of sensitive instruments and an appropriate 
data processing strategy.   

\section{Acknowlegements}

I appreciate the help of M. V. Prokhorov in preparation of the figures.

\end{document}